\pdfoutput=1
\documentclass[aps,pra,twocolumn,showpacs,preprintnumbers,amsmath,amssymb,floatfix]{revtex4-1}
\usepackage{graphicx}
\usepackage[usenames,dvipsnames]{xcolor}
\usepackage{comment}
\usepackage[breaklinks=true]{hyperref}
\usepackage{cleveref}
\usepackage{makecell}
\usepackage{float}
\usepackage{lineno}
\usepackage[T1]{fontenc}
\usepackage[caption=false]{subfig}
\newcommand{\bra}[1]{\ensuremath{\langle #1 |}}
\newcommand{\ket}[1]{\ensuremath{| #1 \rangle}}

\usepackage{amssymb,amsmath} 
\usepackage{amsfonts}
\usepackage{color}
\usepackage[utf8]{inputenc}
\usepackage{lmodern}
\usepackage{epsfig}
\usepackage[normalem]{ulem}
\usepackage{blindtext}
\begin{document}
\title{Luttinger liquid parameters in one-dimensional Rydberg arrays}

\author{Shu-Ao Liao}
\author{Li-Ping Yang}
\email{liping2012@cqu.edu.cn}
\author{Jin Zhang}
\email{jzhang91@cqu.edu.cn}
\affiliation{Department of Physics and Chongqing Key Laboratory for Strongly Coupled Physics, Chongqing University, Chongqing 401331, People's Republic of China}
\definecolor{burnt}{cmyk}{0.2,0.8,1,0}
\def\lt{\lambda ^t}
\def\note{note}
\def\beq{\begin{equation}}
\def\enq{\end{equation}}

\date{\today}
\begin{abstract}
We investigate Berezinskii-Kosterlitz-Thouless (BKT) transitions in one-dimensional Rydberg chains, where commensurate critical regimes associated with the melting of crystalline orders with period larger than five and incommensurate floating phases are both described by Luttinger liquid theory. The central quantity is the Luttinger liquid parameter $K$, which characterizes the universal low-energy theory and controls the relevance of perturbations driving BKT transitions. We extract $K$ using Friedel oscillations and the recently developed crosscap method introduced in Phys. Rev. Lett. 134, 076501 (2025). As benchmarks, we first apply the crosscap method to a $\mathbb{Z}_3$ dual hard-core boson chain and a spin-1 XY chain with single-ion anisotropy, obtaining BKT transition points consistent with previous results after finite-size extrapolation. We then compute $K$ in the Rydberg chain along lines with fixed correlation-oscillation period near BKT transitions. Along the commensurate period-five line, the critical values of $K$ predicted by sine-Gordon theory reveal two BKT transitions separating the disordered phase, the critical phase, and the $\mathbb{Z}_5$ crystalline phase. The results from Friedel oscillations and the crosscap method agree with each other and are further supported by energy-gap scaling and Binder-cumulant analysis. To obtain reliable values of $K$ from Friedel oscillations, we use a multi-harmonic fitting scheme throughout the analysis. Along incommensurate lines, the BKT points obtained from Friedel oscillations agree with those extracted from entanglement entropy. Finally, we show that the values of $K$ obtained from the two methods are mutually consistent inside the incommensurate floating phase.
\end{abstract}


\maketitle

\section{Introduction}\label{sec:introduction}

The Tomonaga-Luttinger liquid (TLL) framework \cite{F.D.M.Haldane_1981,PhysRevB.84.085114} provides a universal description of a broad class of one-dimensional (1D) strongly correlated systems. In such low-dimensional settings, the quasiparticle picture of Landau Fermi liquid theory breaks down, and the low-energy excitations are instead described by collective density fluctuations \cite{giamarchi2004quantum,tsvelik2007quantum,sachdev2001quantum}. Through bosonization, interacting 1D systems can be mapped to an effective theory of free compact bosons. The universal low-energy physics is then governed by two parameters: the sound velocity $v$ and the dimensionless Luttinger parameter $K$. The latter fixes the scaling dimensions of perturbations and controls the algebraic exponents of correlation functions, making it a key quantity for characterizing 1D critical phases and Berezinskii-Kosterlitz-Thouless (BKT) transitions \cite{Berezinsky:1970fr,Kosterlitz_1973,Kosterlitz_1974}.

Accurate extraction of the Luttinger parameter $K$ is therefore essential, but it remains numerically challenging. High-precision simulations of 1D quantum systems are enabled by the density-matrix renormalization group (DMRG) algorithm \cite{PhysRevLett.69.2863,PhysRevB.48.10345}. Within DMRG, $K$ can be obtained from correlation functions and Friedel oscillations in open chains \cite{PhysRevB.65.165122}. It can also be extracted from the oscillating part of the entanglement entropy \cite{CalabreseParityprl2010,XavierRenyiParity2011} or from finite-size excitation spectra within conformal-field-theory analyses \cite{NomuraCritical1994JPAMG,KitazawaTwisted1997JPAMG}, although these approaches often require careful treatment of finite-size corrections. Recently, a complementary wave-function-based approach was introduced in Ref.~\cite{PhysRevLett.134.076501}. This method extracts $K$ from a crosscap overlap of a single finite-size wave function and provides an efficient route to the Luttinger parameter in systems of moderate size.

A prominent realization of TLL physics occurs in the quantum melting of commensurate crystalline phases \cite{giamarchi2004quantum,Pokrovsky1979,PBak_1982,Villain&Bak1981IsingCompeting,HuseFisherPRL1982,Huse1984commensurate,PhysRevB.69.075106}. In many 1D systems, the melting of a commensurate ordered state does not necessarily proceed directly into a disordered phase. Instead, an intermediate gapless critical regime, often referred to as a floating phase, can appear between the ordered and disordered phases \cite{Pokrovsky1979,PBak_1982}. In such a phase, density correlations decay algebraically and oscillate with a wave vector that can vary continuously with microscopic parameters. Although floating phases have a long history in low-dimensional physics, they have recently attracted renewed attention through theoretical studies of effective hard-boson models \cite{PhysRevB.69.075106} and through the development of Rydberg quantum simulators.

Neutral atoms trapped in optical tweezers and excited to Rydberg states realize controllable long-range interacting quantum systems \cite{Labuhn2016RydIsing,Bernien2017Dynamics,Keesling2019Kibble,Leseleuc2019topo,Semeghini2021SL,Ebadi2021_256,Pascal2021AF,ChenContinuous2023,Daniel2024RydStringBreaking}. In 1D Rydberg chains, the blockade mechanism, characterized by the blockade radius $R_b$ in units of the lattice spacing $a$ \cite{Wucpb2021}, strongly constrains nearby Rydberg excitations. Together with the competition among the Rabi frequency $\Omega$, the detuning $\Delta$, and the repulsive van der Waals interaction $C_6/r^6$, this blockade physics gives rise to a rich phase diagram with disordered phases, commensurate crystalline orders, and intermediate floating phases \cite{PhysRevLett.122.017205,Chepiga&Mila2021Kibble,PhysRevResearch.4.043102,zhang_probing_2025,Soto-GarciaNumerical2025PRR,PhysRevB.111.165154}. A useful way to organize this phase diagram is to track the dominant density-correlation wave vector $k$, or equivalently the oscillation period $p=2\pi/k$ in units of the lattice spacing. This period characterizes short-range density oscillations in the disordered phase, long-range density order in the crystalline phase, and algebraic oscillations in the floating phase. Following lines of fixed $p$ therefore provides a natural route to compare different regimes of the phase diagram and to analyze how the Luttinger parameter evolves near BKT transitions. Recent experiments have also demonstrated TLL behavior in a Rydberg-encoded dipolar XY spin chain \cite{qfnp-6dpz}, highlighting the broader potential of Rydberg platforms for probing one-dimensional critical phenomena.

The role of commensurability and topological defects in melting problems was analyzed by Huse and Fisher \cite{HuseFisherPRL1982,Huse1984commensurate}. They showed that, depending on the commensurability and on the relevance of domain-wall and dislocation excitations, several scenarios may occur between a commensurate ordered phase and a disordered phase, including direct chiral transitions and two-step melting through an intermediate floating phase. Related two-stage melting of commensurate Rydberg excitation solids has also been discussed in Rydberg atom arrays \cite{Weimer2010twostage}. In the usual Rydberg-chain phase diagram defined by $R_b/a$ and $\Delta/\Omega$, the floating phase is expected to be separated from the commensurate crystalline order by a Pokrovsky-Talapov (PT) transition \cite{Pokrovsky1979} and from the disordered fluid phase by a BKT transition.

The commensurate period-five case is especially subtle. For clock-type models with $\mathbb{Z}_p$ symmetry, an emergent $U(1)$ description appears for $p\geq 5$, leading to BKT physics and an intermediate critical phase \cite{ORTIZ2012780,PhysRevB.100.094428,li2020}. This raises a natural question for the Rydberg chain: along the constant $p=5$ line, is the transition between the floating regime and the $\mathbb{Z}_5$ crystalline order still of PT type, as expected from the usual commensurate-incommensurate melting scenario, or does it instead become BKT-like because of the emergent $U(1)$ physics associated with the $\mathbb{Z}_5$ symmetry? To our knowledge, this question has not been resolved in previous studies. More generally, despite substantial progress in mapping the global phase diagram of 1D Rydberg chains \cite{rader2019floating,PhysRevResearch.4.043102}, a quantitative characterization of the corresponding critical behavior and Luttinger parameters remains incomplete.

In this paper, we employ DMRG to determine the Luttinger parameter in 1D Rydberg chains, both deep inside the floating phase and near BKT transitions. We combine Friedel-oscillation fitting, entanglement-entropy scaling, rescaled energy-gap analysis, Binder-cumulant analysis, and the crosscap method. We first benchmark the crosscap method on two well-understood nonintegrable models: the spin-1 XY model with single-ion anisotropy, which is equivalent to the O(2) model \cite{LutherCritical1977PRB,ZouProgress2014PRA,PhysRevB.103.245137}, and the $\mathbb{Z}_3$ dual hard-core boson model \cite{WhitsittQuantum2018PRB,PhysRevResearch.3.023049}. In both cases, the critical points determined by the crossing of the extracted $K$ with the theoretical critical value $K_c$, followed by extrapolation to the thermodynamic limit, agree with established results.

We then apply these methods to the 1D Rydberg chain. Along the commensurate $p=5$ line, our calculations reveal that the system undergoes two distinct BKT transitions as it evolves from the disordered phase to the $\mathbb{Z}_5$ ordered phase. These two critical points are very close to each other, so that the standard Binder-cumulant analysis shows only a single crossing and fails to resolve them separately. In contrast, the Luttinger parameter extracted from both the crosscap method and Friedel-oscillation fitting intersects the theoretical critical values $K_c$ at two distinct locations. The small separation between the two BKT transition points is further supported by the rescaled energy gap, which also displays two distinct crossings. This behavior closely resembles the finite-size scaling structure of the $p=5$ quantum clock model \cite{PhysRevLett.134.076501}. The first transition is associated with the boundary between the disordered and critical phases, while the second transition separates the critical phase from the period-five density wave.

Finally, we extend our analysis to incommensurate regimes. We calculate the Luttinger parameter along representative lines with $p=3.4$, located between the $\mathbb{Z}_3$ and $\mathbb{Z}_4$ crystalline orders, and $p=4.5$, located between the $\mathbb{Z}_4$ and $\mathbb{Z}_5$ crystalline orders \cite{incommen-analysis}. In these regimes, the critical points determined by the crossing of $K$ from Friedel oscillations with the corresponding $K_c$ agree with the locations of entanglement-entropy peaks. Inside the incommensurate floating phase, the values of $K$ obtained from Friedel oscillations and from the crosscap method are mutually consistent, demonstrating that the two approaches provide complementary probes of TLL behavior in Rydberg chains.

This paper is organized as follows. In Sec.~\ref{sec:model}, we introduce the Rydberg-chain Hamiltonian and the methods for extracting the Luttinger parameter $K$, including Friedel-oscillation fitting and the crosscap method. In Secs.~\ref{subsec:BenchmarkZ3} and \ref{subsec:BenchmarkXY}, we benchmark the crosscap method using the $\mathbb{Z}_3$ dual hard-core boson chain and the spin-1 XY chain with single-ion anisotropy. In Sec.~\ref{subsec:chain}, we present the global phase diagram of the 1D Rydberg chain and study phase transitions along one commensurate and two incommensurate constant-$k$ trajectories. In Secs.~\ref{subsec:commensurate} and \ref{subsec:incommensurate}, we determine the BKT transition points along these trajectories using $K$, the rescaled energy gap, Binder cumulant, and entanglement entropy.  Finally, we summarize our results and discuss future directions in Sec.~\ref{sec:conclusion}.

\section{Model and Methods}\label{sec:model}

\subsection{Model Hamiltonian} \label{subsec:modelHam}

The dynamics of a one-dimensional chain of Rydberg atoms is described by the Hamiltonian
\begin{equation} 
\hat{H} = \sum_{i=1}^{N} \left[\frac{\Omega}{2}  \left(\ket{g_i}\bra{r_i} + \text{H.c.}\right) - \Delta \hat{n}_i \right] + \sum_{i < j} V_{ij} \hat{n}_i\hat{n}_{j}. 
\end{equation}
Here, $\ket{g_i}$ and $\ket{r_i}$ denote the ground and Rydberg states of the atom at site $i$, respectively, and $\hat{n}_i=\ket{r_i}\bra{r_i}$ is the Rydberg excitation number operator. The parameter $\Omega$ is the Rabi frequency of the coherent coupling between $\ket{g_i}$ and $\ket{r_i}$, and $\Delta$ is the laser detuning from the atomic resonance. In experiments, this transition is often driven through a two-photon process \cite{Bernien2017Dynamics}. The system size is denoted by $N$. The interaction between Rydberg excitations is repulsive and takes the van der Waals form $V_{ij}=C_6/(|i-j|a)^6$, where $a$ is the lattice spacing and $C_6$ is the van der Waals coefficient. It is useful to parameterize the interaction strength by the blockade radius $R_b$, defined by $V(R_b)=\Omega$. With this definition, the interaction can be written as $V_{ij}=\Omega R_b^6/(|i-j|a)^6$. The dimensionless ratio $R_b/a$ controls the extent of the blockade constraint relative to the lattice spacing. Together with the detuning-to-Rabi-frequency ratio $\Delta/\Omega$, it determines the competition between coherent driving, detuning, and repulsive interactions, leading to a variety of spatially ordered Rydberg-crystal phases and intervening critical regimes \cite{rader2019floating,PhysRevResearch.4.043102}.

\subsection{Friedel oscillations} \label{subsec:Friedel}

For a finite chain with open boundary conditions (OBCs), the boundaries induce Friedel oscillations in the local density. Boundary conformal field theory (CFT) gives the leading density modulation as \cite{SchulzPRL1990,PhysRevB.51.17827}
\begin{equation}\label{eq:Frideloscillation}
\left< n_{j}\right> - \overline{n} \propto
\frac{\cos(qj)}{\left[(N/\pi)\sin(\pi j /N)\right]^{K}},
\end{equation}
where $q$ is the wave vector of the density oscillation, $N$ is the system size, and $\overline{n}$ is the average density. The Luttinger parameter $K$ can therefore be extracted by fitting the spatial decay of the Friedel oscillations in the local occupation $\left< n_j \right>$ \cite{PhysRevResearch.4.043225}.

In practice, fitting only the leading harmonic can lead to noticeable deviations because higher harmonics may have non-negligible contributions \cite{arXiv:2604.24889}. To account for these corrections, we use a multi-harmonic fitting ansatz for the local density \cite{giamarchi2004quantum,PhysRevB.51.17827,PhysRevResearch.3.023049},
\begin{equation}\label{eq:harmonicFrideloscillation}
\langle n_{j} \rangle = \bar{n} + \sum_{m=1}^{M} A_m
\frac{\cos(m q j + \phi_m)}
{\left[(N/\pi)\sin(\pi j /N)\right]^{m^2 K}},
\end{equation}
where $A_m$ and $\phi_m$ are the amplitude and phase shift of the $m$th harmonic, respectively. In our analysis, we typically include harmonics up to $m=3$. This multi-harmonic fitting procedure reduces the influence of subleading oscillatory components and yields a more stable estimate of $K$ from the decay exponent of the leading Friedel oscillation.

\subsection{Crosscap method} \label{subsec:crosscapMethod}

The Luttinger parameter $K$ controls the universal low-energy properties of a one-component TLL. Since a TLL is described by a compactified-boson CFT, $K$ can be extracted from universal properties of the many-body wave function. Ref.~\cite{PhysRevLett.134.076501} proposed the eigenstate crosscap method, which determines $K$ from a single finite-size wave function with periodic boundary conditions (PBCs). The key idea is to impose a crosscap contraction on the ground-state wave function: for an even-length periodic chain, each site $i$ is paired with the antipodal site $i+N/2$ through a maximally entangled state in the local basis.

Formally, this contraction is equivalent to evaluating the overlap between the ground-state wave function $|\Psi\rangle$ and a lattice crosscap state $|\mathcal{C}_{\rm latt}\rangle$. For the Rydberg chain, the local basis is ${|g\rangle,|r\rangle}$, and the antipodal pair is contracted with the maximally entangled state $\left(|g_i g_{i+N/2}\rangle+|r_i r_{i+N/2}\rangle\right)/\sqrt{2}$. In our MPS implementation, we do not explicitly construct $|\mathcal{C}_{\rm latt}\rangle$ as a separate many-body state. Instead, we directly contract the MPS tensors according to this antipodal pairing pattern. In the TLL regime, the resulting crosscap overlap approaches the universal value
\begin{equation}\label{crosscapEq}
\left|\left< \mathcal{C}_{\rm latt} | \Psi \right>\right|^{2}
=
\frac{1}{\sqrt{K}} .
\end{equation}
Thus, $K$ can be obtained directly from the crosscap contraction of one ground-state wave function. This wave-function-based estimator provides an independent probe of $K$. Unlike Friedel-oscillation fitting, it does not require choosing a fitting window or resolving the spatial decay of oscillatory correlations. Finite-size corrections are still present and can become substantial near phase boundaries, so we use the crosscap method and Friedel-oscillation fitting as complementary diagnostics.

In this work, we mainly use Friedel oscillations and the crosscap method to determine $K$. We also use the entanglement entropy \cite{RevModPhys.82.277}, structure factor, Binder cumulant \cite{BINDER19851}, and rescaled energy gap \cite{MishraPhase2011PRB} to identify phases and locate phase boundaries. These diagnostics have been described in detail in our recent works \cite{zhang_probing_2025,PhysRevB.111.165154} and are used here as supporting tools for a systematic analysis of the Rydberg chain.

\subsection{Parameters of DMRG algorithms} \label{subsec:paraDMRG}

We compute the ground states using the finite-size density-matrix renormalization group (DMRG) algorithm formulated in the matrix-product-state (MPS) framework \cite{PhysRevLett.75.3537}, implemented with the \textsc{ITensor Julia Library} \cite{10.21468/SciPostPhysCodeb.4}. For the Rydberg chain, we keep van der Waals interactions up to a distance of 20 lattice sites, which is sufficient to capture the relevant interaction range in our simulations. For calculations with OBCs, the maximum bond dimension is increased until the truncation error is below $10^{-11}$. For calculations of the crosscap contraction, we use PBCs and keep the truncation error below $10^{-10}$. Other truncation-error settings, when used, are specified in the corresponding sections. Convergence is checked by monitoring both the ground-state energy and the entanglement entropy, and the sweeps are stopped when the absolute changes between two successive sweeps are smaller than $10^{-11}$ for the energy and $10^{-8}$ for the entanglement entropy. The system sizes are chosen to be compatible with the spatial periodicity of the crystalline ordered phases or with the dominant wave vector $k$ in the floating phase. Typically, tens of sweeps are sufficient to reach convergence deep in the disordered and crystalline phases, whereas up to thousands of sweeps are required in the floating phase and near BKT transition points.

\section{Results} \label{sec:results}

Prior to analyzing the Rydberg chain, we first benchmark the crosscap-overlap method and assess its numerical accuracy. We apply it to two well-established nonintegrable 1D models with known BKT transitions: the $\mathbb{Z}_3$ dual hard-core boson model and the spin-1 XY chain with single-ion anisotropy, which realizes the quantum $O(2)$ universality class. These benchmark examples provide controlled test cases for validating the extraction of the Luttinger parameter $K$ before applying the method to the Rydberg chain.

\begin{figure}[htp]
\centering    
\includegraphics[width=1\linewidth]{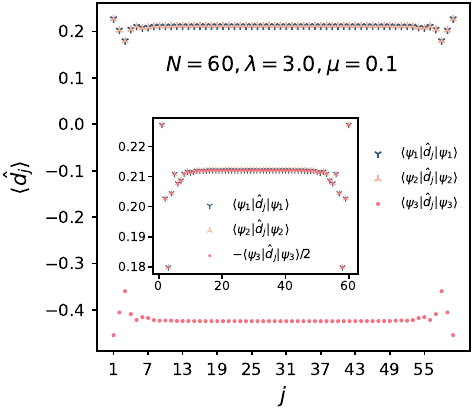}
\caption{Local profiles of $\langle \hat{d}_j \rangle$ for the three degenerate ground states $|\psi_1\rangle$, $|\psi_2\rangle$, and $|\psi_3\rangle$ deep in the $\mathbb{Z}_3$-ordered phase, obtained for $N=60$ as a function of the site index $j$. The expectation values of $\hat{d}_j$ in $|\psi_1\rangle$ and $|\psi_2\rangle$ are degenerate and have half the magnitude of that in $|\psi_3\rangle$. The inset illustrates this relation by plotting the rescaled profile $-\langle \psi_3| \hat{d}_j | \psi_3\rangle/2$, which collapses onto the profiles of $|\psi_1\rangle$ and $|\psi_2\rangle$.}
\label{fig:Z3dualhardcoreDisityProfile}
\end{figure}

\begin{figure}[htp]
\centering    
\includegraphics[width=0.95\linewidth]{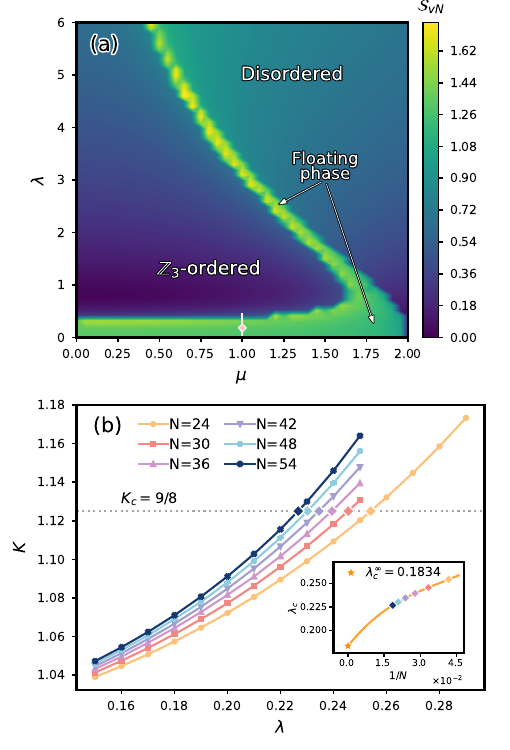}
\caption{(a) Ground-state phase diagram of the $\mathbb{Z}_3$ dual model for the quantum phase transitions out of the period-3 ordered phase of the Rydberg chain. The phase diagram is mapped out using the von Neumann entanglement entropy $\mathcal{S}_{\rm vN}$. The data are obtained for $N=480$ with OBCs, and $\mathcal{S}_{\rm vN}$ is evaluated at the middle bond, between sites 240 and 241. The highlighted wedge-shaped region denotes the intermediate floating phase, which separates the $\mathbb{Z}_3$-ordered phase at smaller $\mu$ and $\lambda$ from the disordered phase at larger $\mu$ and $\lambda$. (b) Luttinger 
$K$ extracted from the crosscap overlap as a function of $\lambda$ at fixed $\mu=1$ for different system sizes $N$. The horizontal line marks the theoretical BKT critical value $K_c=9/8$. Its intersections with the $K_N(\lambda)$ curves, indicated by colored diamonds, determine the finite-size critical points $\lambda_c(N)$. The inset shows the finite-size extrapolation of $\lambda_c(N)$ to the thermodynamic limit, with the extrapolated critical point marked by a pink star. The fitting function is $f(1/N)=0.1834+3.2326(1/N)-57.5961(1/N)^2+497.5602(1/N)^3$.}
\label{fig:Z3dualPhaseDiagram&crosscap}
\end{figure}

\subsection{$\mathbb{Z}_3$ dual hard-core boson model} \label{subsec:BenchmarkZ3}

We first benchmark the crosscap-overlap method using the $\mathbb{Z}_3$ dual hard-core boson model, which provides an effective dual description of the quantum phase transition out of the period-3 ordered phase of the Rydberg chain \cite{PhysRevResearch.3.023049}. The Hamiltonian is
\begin{eqnarray}\label{eq:dualp-3rydbergham}
\nonumber \hat{H} &=& \sum_{i=1}^{N} \left[-t\left(\hat{d}_{i}^{\dagger}\hat{d}_{i+1}+\text{H.c.}\right) - \mu \hat{n}_{i}\right] \\
&+& \sum_{i=1}^{N}\lambda \left(\hat{d}_{i}^{\dagger}\hat{d}_{i+1}^{\dagger}\hat{d}_{i+2}^{\dagger}+\text{H.c.}\right),
\end{eqnarray}
where $\hat{d}_i$ is a hard-core boson annihilation operator and $\hat{n}_i=\hat{d}_i^\dagger \hat{d}_i$. The three-particle creation and annihilation term breaks the usual $U(1)$ particle-number conservation down to a $\mathbb{Z}_3$ conservation law, so that the total occupation number is conserved modulo three. The corresponding $\mathbb{Z}_3$ symmetry operation is generated by $\hat{Q}=\omega^{\hat{N}}$, where $\hat{N}=\sum_i \hat{d}_i^\dagger \hat{d}_i$ and $\omega=e^{i2\pi/3}$. Under this operation, the hard-core boson operators transform as $\hat{Q}\hat{d}_i\hat{Q}^{-1}=\omega^* \hat{d}_i$ and $\hat{Q}\hat{d}_i^\dagger\hat{Q}^{-1}=\omega \hat{d}_i^\dagger$.

In the $\mathbb{Z}_3$-ordered phase, the ground states become threefold degenerate in the thermodynamic limit. We label the three lowest states in the $\mathbb{Z}_3$ charge sectors by $|0\rangle$, $|1\rangle$, and $|2\rangle$, corresponding to total occupation numbers equal to $0$, $1$, and $2$ modulo $3$, respectively. These fixed-$\mathbb{Z}_3$-charge states are eigenstates of $\hat{Q}$ and therefore preserve the $\mathbb{Z}_3$ symmetry. The symmetry-broken states, which have definite values of the order parameter, are obtained as their linear combinations,
\begin{eqnarray}
|\psi_1\rangle &=& \frac{1}{\sqrt{3}}\left(|0\rangle+\omega |1\rangle+\omega^* |2\rangle\right), \nonumber\\
|\psi_2\rangle &=& \frac{1}{\sqrt{3}}\left(|0\rangle+\omega^* |1\rangle+\omega |2\rangle\right), \nonumber\\
|\psi_3\rangle &=& \frac{1}{\sqrt{3}}\left(|0\rangle+|1\rangle+|2\rangle\right).
\end{eqnarray}
In these symmetry-broken states, the local order parameter $\langle \hat{d}_j\rangle$ takes three symmetry-related values that differ by phases $1$, $\omega$, and $\omega^*$. If $\langle \psi_3|\hat{d}_j|\psi_3\rangle$ is chosen to be real, then the other two order parameters are complex conjugates whose real parts are equal to $-\langle \psi_3|\hat{d}_j|\psi_3\rangle/2$.

Our DMRG calculations are performed in a real basis. Therefore, the two complex-conjugate symmetry-broken states cannot appear as two separate complex states in the numerical calculation. Instead, DMRG returns real linear combinations of them, forming a real doublet. Consequently, when one numerically obtained state has the real order-parameter profile $\langle \psi_3|\hat{d}_j|\psi_3\rangle$, the other two states in the real doublet have expectation values of $\hat{d}_j$ corresponding to the real parts of the two complex-conjugate order parameters. This explains why their profiles are equal to $-\langle \psi_3|\hat{d}_j|\psi_3\rangle/2$. As shown in Fig.~\ref{fig:Z3dualhardcoreDisityProfile}, the rescaled profile $-\langle \psi_3|\hat{d}_j|\psi_3\rangle/2$ collapses onto the profiles of the real doublet, confirming the expected $\mathbb{Z}_3$ structure of the ordered ground states. 

The model exhibits a BKT transition between a critical floating phase and a gapped $\mathbb{Z}_3$-ordered phase. We focus on the vertical cut at $\mu=1$, shown in Fig.~\ref{fig:Z3dualPhaseDiagram&crosscap}(a), and compute the Luttinger parameter $K$ for different system sizes using the crosscap-overlap method. The intersections between the $K_N(\lambda)$ curves and the theoretical critical value $K_c=9/8$ \cite{giamarchi2004quantum} determine the finite-size critical couplings $\lambda_c(N)$. As shown in the inset of Fig.~\ref{fig:Z3dualPhaseDiagram&crosscap}(b), we extrapolate $\lambda_c(N)$ as a function of $1/N$ using a third-degree polynomial and obtain the thermodynamic critical point $\lambda_c \approx 0.1834$. This value agrees well with the estimate $\lambda_c \approx 0.187$ reported in Ref.~\cite{PhysRevResearch.3.023049}, where the transition point was determined from the crossing of the Friedel-oscillation estimate of $K$ with $K_c=9/8$. In the same work, the BKT nature of the transition was further supported by the correlation-length scaling in the ordered phase, $\ln \xi \sim 1/\sqrt{\lambda-\lambda_c}$. The extrapolated critical point is marked by a diamond on the vertical cut in Fig.~\ref{fig:Z3dualPhaseDiagram&crosscap}(a).

\begin{figure}[htp]
\centering    
\includegraphics[width=1\linewidth]{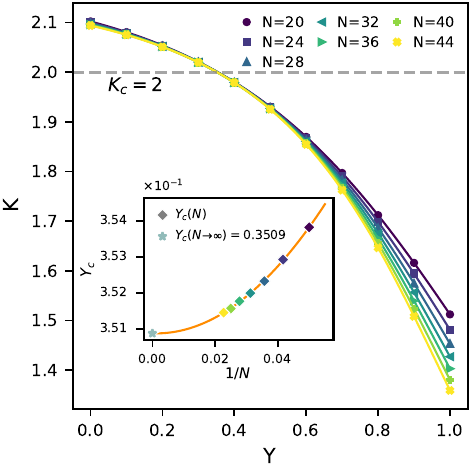}
\caption{The Luttinger parameter $K_N(Y)$ for the spin-1 XY chain with single-ion anisotropy, extracted from the crosscap overlap as a function of $Y$ at fixed $X=1$ for different system sizes $N$. The dashed line marks the theoretical critical value $K_c=2$. The crossings between $K_N(Y)$ and $K_c=2$ determine the finite-size critical points $Y_c(N)$. The inset shows the polynomial extrapolation of $Y_c(N)$ to the thermodynamic limit. The extrapolated value, indicated by the pink star, is $Y_c=0.3509$. The fitting function is $f(1/N)=1.2369(1/N)^2-0.0030(1/N)+0.3509$.} 
\label{fig:O2S=1crosscap}
\end{figure}
\subsection{Spin-1 XY chain with single-ion anisotropy} \label{subsec:BenchmarkXY}
 
As a second benchmark, we consider the one-dimensional quantum $O(2)$ model in the charge representation \cite{PhysRevB.103.245137},
\begin{equation}
     \hat{H}_{c}=\frac{Y}{2} \sum_{l=1}^{L}\left(\hat{S}_{l}^{z}\right)^{2}
     -\frac{X}{2} \sum_{l=1}^{L-1}\left(\hat{U}_{l}^{+} \hat{U}_{l+1}^{-}
     +\hat{U}_{l}^{-} \hat{U}_{l+1}^{+}\right).
\end{equation}
Here, $X$ and $Y$ are effective coupling parameters. In the charge basis, the operators satisfy
$\hat{S}^{z}\ket{n}=n\ket{n}$ and $\hat{U}^{\pm}\ket{n}=\ket{n\pm1}$.
In the spin-$S$ truncation, the local Hilbert space is restricted to $|n|\leq S$. In this work, we focus on the $S=1$ truncation, for which the model is mapped to a spin-1 XY chain with single-ion anisotropy.

Figure~\ref{fig:O2S=1crosscap} shows the Luttinger parameter $K_N(Y)$ extracted from the crosscap overlap at fixed $X=1$ for system sizes from $N=20$ to $N=44$ with interval $4$. The theoretical BKT critical value is $K_c=2$ \cite{giamarchi2004quantum}. The crossings between $K_N(Y)$ and $K_c=2$ determine the finite-size critical points $Y_c(N)$. The inset of Fig.~\ref{fig:O2S=1crosscap} shows the extrapolation of $Y_c(N)$ to the thermodynamic limit using a second-degree polynomial in $1/N$, $f(1/N)=1.2369(1/N)^2-0.0030(1/N)+0.3509$. This gives $Y_c=0.3509$, in good agreement with previous estimates in Ref.~\cite{PhysRevB.103.245137}, where $Y_c=0.3507$ was obtained from level spectroscopy and $Y_c=0.3512$ from gap scaling.

It is worth noting that the finite-size crossings obtained from the crosscap overlap are already close to the thermodynamic values. Even for the smallest size used here, $N=20$, the crossing point is below $Y=0.354$, differing only slightly from the extrapolated value. A similar trend is found in the $\mathbb{Z}_3$ dual hard-core boson model discussed above, where the smallest system size $N=24$ gives $\lambda_c(N=24)=0.25$, still reasonably close to the thermodynamic value $\lambda_c=0.1834$. In contrast, conventional probes of BKT transitions often suffer from strong logarithmic finite-size corrections, so that the apparent finite-size transition points can be far from the thermodynamic limit and large systems with careful scaling analyses are required. For example, entanglement entropy has also been used to locate the same BKT transition in this spin-1 XY chain \cite{ZhangFidelity2021PRB}. In that analysis, systems with more than 1000 sites were included in the fitting, yielding $Y_c=0.353$. The consistency between the crosscap-overlap results and established estimates in both benchmark models supports the use of the crosscap method as an efficient and reliable estimator of $K$ before applying it to the Rydberg chain.

\begin{figure}[htp]
\centering    
\includegraphics[width=1\linewidth]{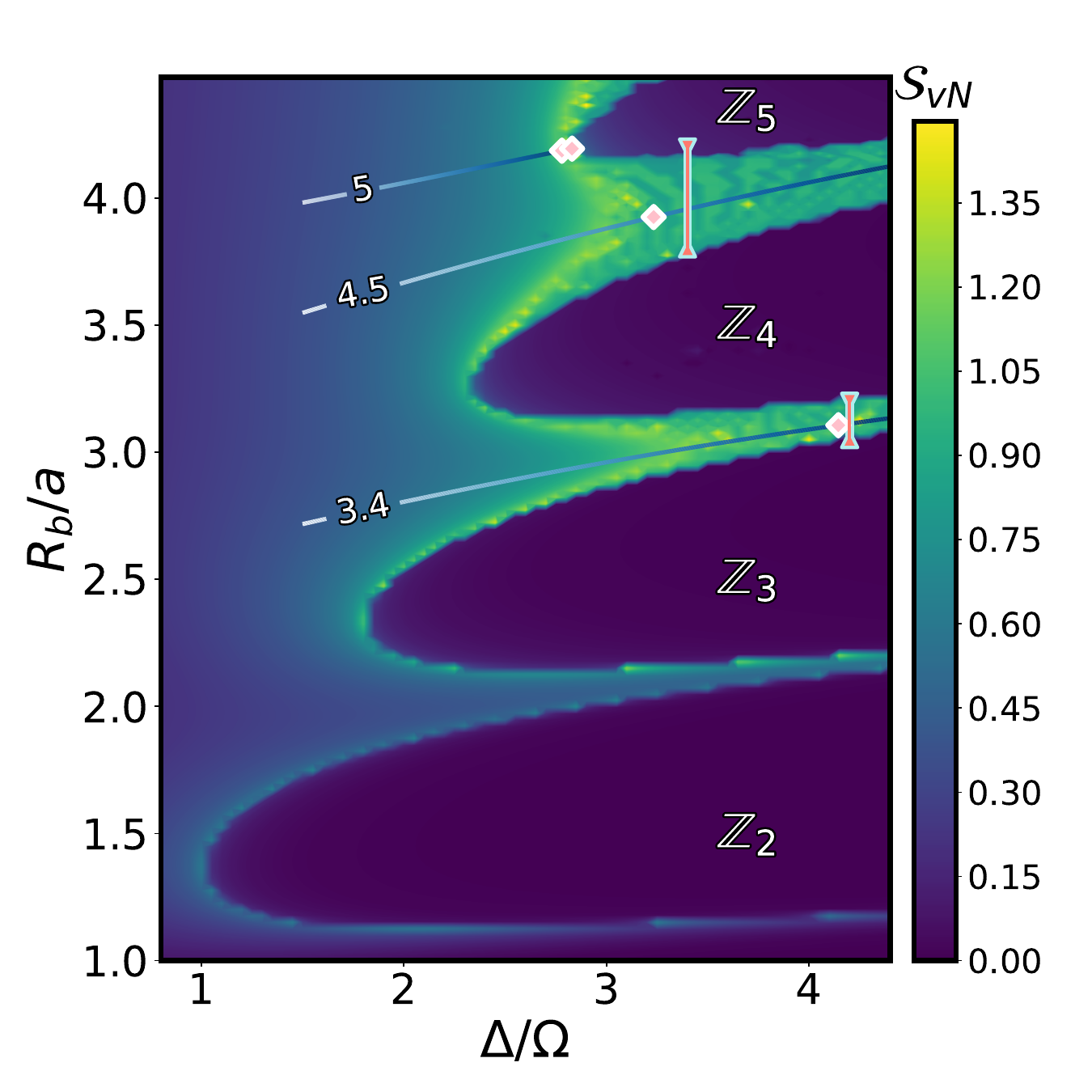}
\caption{Phase diagram of the one-dimensional Rydberg chain, mapped out using the von Neumann entanglement entropy $\mathcal{S}_{\rm vN}$. The numerical results are obtained for an open chain with $N=721$ sites, and $\mathcal{S}_{\rm vN}$ is evaluated at the middle bond, between sites 360 and 361. The dark lobes correspond to the $\mathbb{Z}_p$ crystalline ordered phases, where $p$ denotes the characteristic spatial period. Intermediate floating phases are observed between the $\mathbb{Z}_3$ and $\mathbb{Z}_4$ phases, between the $\mathbb{Z}_4$ and $\mathbb{Z}_5$ phases, and surrounding the $\mathbb{Z}_5$ phase on the side facing the disordered regime. The solid blue-gradient curves represent lines of constant dominant wave vector $k$, labeled by the corresponding spatial period $p=2\pi/k$ with $p=3.4$, $4.5$, and $5$. These constant-$p$ lines indicate the trajectories used in the following numerical analyses. The BKT transition points are marked by diamonds. Along the constant-$p=5$ line, two nearby BKT transitions are resolved. Two vertical cuts in representative floating regions are also indicated for later discussion.}
\label{fig:1DPhaseDiagramwithp3.4linep4.5line&p5line}
\end{figure}

\subsection{BKT transitions in the Rydberg chain} \label{subsec:chain}

Figure~\ref{fig:1DPhaseDiagramwithp3.4linep4.5line&p5line} shows the phase diagram of the one-dimensional Rydberg chain obtained for an open chain with $N=721$ sites. The color intensity represents the von Neumann entanglement entropy $\mathcal{S}_{\rm vN}$ evaluated at the central bond, between sites 360 and 361. The dark lobe-shaped regions correspond to commensurate $\mathbb{Z}_p$ crystalline phases, where $p=2\pi/k$ is the characteristic spatial period and $k$ is the dominant density-wave vector. Inside each crystalline lobe, the density wave is locked to a commensurate value of $p$. Outside the lobes, the same commensurate values continue only along constant-$k$ trajectories, which connect to the corresponding crystalline regions and extend toward smaller $\Delta/\Omega$. Between neighboring commensurate trajectories, the dominant wave vector becomes incommensurate and varies continuously in the thermodynamic limit. In a finite chain, this continuous variation appears as a sequence of discrete plateaus.

The melting behavior depends strongly on the commensurability $p$. The $\mathbb{Z}_2$ ordered phase melts through an Ising transition. For the $\mathbb{Z}_3$ and $\mathbb{Z}_4$ ordered phases, several scenarios are possible. Along special commensurate lines, the direct transition is described by CFT with dynamical exponent $z=1$, such as the three-state Potts CFT for $p=3$ and the Ashkin-Teller CFT for $p=4$. Moving away from these commensurate critical points, chiral perturbations can first lead to direct chiral transitions with $1<z<2$. Farther away, in the incommensurate regime, the melting can split into two transitions through an intermediate floating phase \cite{Huse1984commensurate}. In this two-step scenario, the floating phase is bounded by a BKT transition on the disordered side and a Pokrovsky-Talapov transition on the crystalline side \cite{Pokrovsky1979,LazaridesPT2009}. For $p>4$, the $\mathbb{Z}_p$ anisotropy is irrelevant over part of the Gaussian fixed line, allowing an intermediate critical phase with central charge $c=1$ between the disordered and discrete ordered phases \cite{Jose1977,Zamolodchikov1985}. This structure motivates our analysis along constant-$p$ trajectories, including the commensurate $p=5$ line and representative incommensurate lines with $p=3.4$ and $p=4.5$.

Special care is needed near the boundaries between neighboring constant-$k$ plateaus in finite systems. In these crossover regions, the low-energy state can contain sizable contributions from adjacent wave vectors, $k$ and $k+\delta k$, with $\delta k=2\pi/N$. Their interference produces a beat pattern in the density profile,
$\cos(kx)+\cos[(k+\delta k)x]=2\cos[(k+\delta k/2)x]\cos(\delta k x/2)$.
For an open chain, the slowly varying factor $\cos(\delta k x/2)$ can strongly suppress the Friedel-oscillation amplitude near the chain center. Indeed, for $x\simeq N/2$, one has $\delta k x/2\simeq (2\pi/N)(N/2)/2=\pi/2$, giving a node in the beat envelope. If such a beat-modulated profile is fitted by a pure power-law envelope, this suppression can be incorrectly interpreted as a rapid decay, leading to an artificially large value of the extracted Luttinger parameter $K$. Higher harmonics in Eq.~\eqref{eq:harmonicFrideloscillation} can also become important in the small-$K$ regime. Therefore, in the following analysis, we extract critical exponents only along stable constant-$k$ trajectories, avoid the crossover boundaries between adjacent wave-vector plateaus, and use the multi-harmonic fitting ansatz in Eq.~\eqref{eq:harmonicFrideloscillation}.

\subsubsection{Along the commensurate line} \label{subsec:commensurate}

\begin{figure}[htp]
\centering    
\includegraphics[width=1\linewidth]{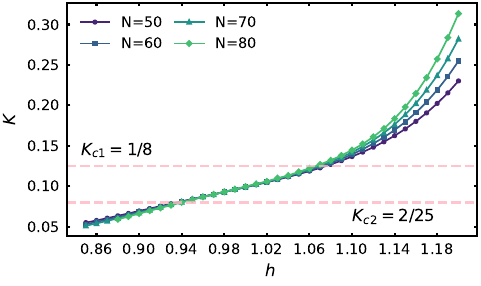}
\caption{Luttinger parameter $K$ of the five-state quantum clock model in Eq.~\eqref{eq:clock_Hamiltonian}, extracted from the crosscap overlap in the $\sigma$ representation and shown as a function of the transverse field $h$. The two dashed lines mark the BKT critical values $K_{c1}=1/8$ and $K_{c2}=2/25$.}
\label{fig:p-clocksigmaBasiscrosscap}
\end{figure}

\begin{figure}[htp]
\centering    
\includegraphics[width=1\linewidth]{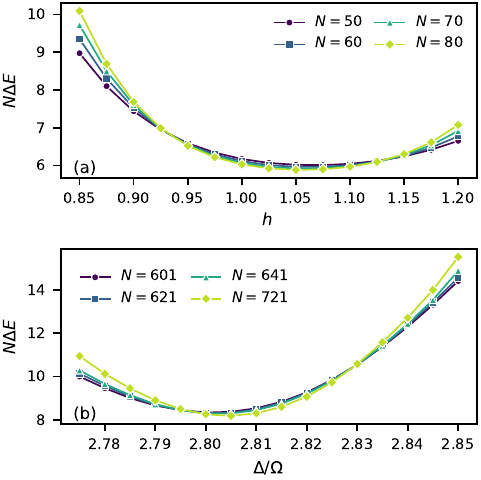}
\caption{Rescaled energy gap $N\Delta E$ for (a) the pinned five-state clock model, defined by Eq.~\eqref{eq:clock_Hamiltonian} with the boundary pinning term in Eq.~\eqref{eq:clock_pinning_term}, and (b) the Rydberg chain along the constant-$p=5$ line. In both models, two crossings appear within the plotted parameter window, signaling two nearby BKT transitions.}
\label{fig:NdEvsDelta-clock5}
\end{figure}

\begin{figure}[htp]
\centering    
\includegraphics[width=1\linewidth]{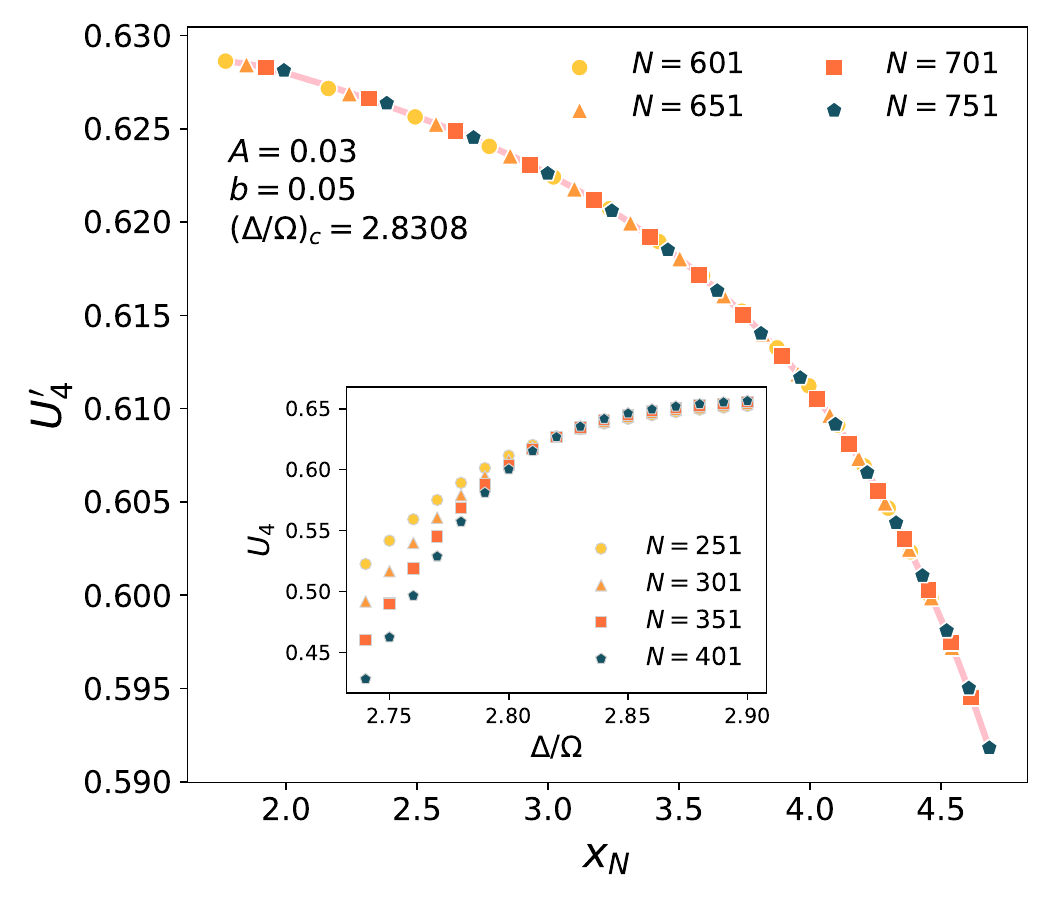}
\caption{Data collapse of the modified Binder cumulant $U_4'=U_4(1+A/\ln N)$ along the constant-$p=5$ line. The horizontal axis is $x_N=\ln N-b/\sqrt{\delta}$, with $\delta=(\Delta/\Omega)_c-\Delta/\Omega$. The best collapse is obtained for $(\Delta/\Omega)_c=2.8308$, $A=0.03$, and $b=0.05$. The inset shows the raw Binder cumulant $U_4$ as a function of $\Delta/\Omega$, where only a single crossing is visible in the plotted range. The collapse function is $f(x)$ with $x=x_N$, where
$f(x)=-3.693\times10^{-5}x^8+0.001064x^7-0.01325x^6+0.09274x^5-0.3985x^4+1.075x^3-1.779x^2+1.646x-0.02102$.}
\label{fig:KTU4forp5line}
\end{figure}

\begin{figure*}[htp]
\centering    
\includegraphics[width=1\linewidth]{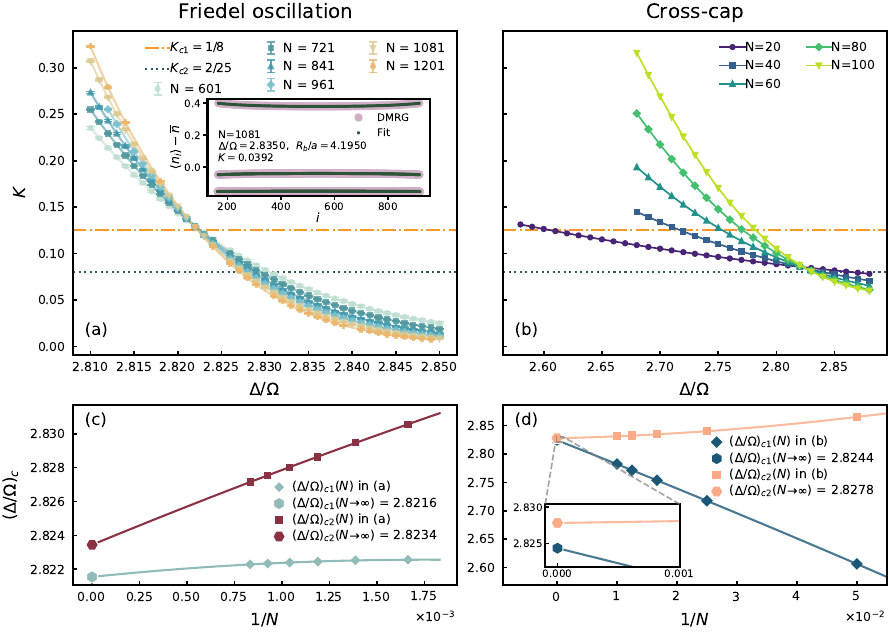}
\caption{Luttinger parameter and finite-size extrapolation of the two BKT transition points along the constant-$p=5$ line. (a) $K$ extracted from Friedel oscillations using the multi-harmonic ansatz in Eq.~\eqref{eq:harmonicFrideloscillation}, with $q=2\pi/5$. The inset compares the DMRG density profile with the Friedel-oscillation fit at $(\Delta/\Omega,R_b/a)=(2.835,4.195)$ for $N=1081$. (b) $K$ extracted from the crosscap overlap for different system sizes. In (a) and (b), the dash-dotted and dotted lines mark the BKT critical values $K_{c1}=1/8$ and $K_{c2}=2/25$, respectively. Their intersections with the finite-size $K$ curves determine the corresponding BKT pseudocritical points. (c) Finite-size extrapolation of the pseudocritical points obtained from Friedel oscillations, giving $(\Delta/\Omega)_{c1}=2.8216$ for the transition between the disordered and floating phases and $(\Delta/\Omega)_{c2}=2.8234$ for the transition between the floating and $\mathbb{Z}_5$ ordered phases. (d) Corresponding extrapolation from the crosscap-overlap data, giving $(\Delta/\Omega)_{c1}=2.8244$ and $(\Delta/\Omega)_{c2}=2.8278$. The inset shows a zoom-in of the extrapolated points. All extrapolations are performed using second-degree polynomial fits in $1/N$.}
\label{fig:1Dp5lineFriedelOscillation&crosscap}
\end{figure*}

We first determine the constant-$p$ trajectories following the procedure described in our previous work \cite{PhysRevB.111.165154}. We perform finite-size DMRG calculations on a grid of parameters $(\Delta/\Omega,R_b/a)$ in the disordered regime at smaller $\Delta/\Omega$. For each point, the dominant wave vector $k$ is extracted from the peak of the structure factor, obtained from the Fourier transform of the ground-state density-density correlation function. We then interpolate the relation between $k$ and the microscopic parameters and determine the parameter values corresponding to the target period $p=2\pi/k$. Finally, the finite-size results are extrapolated to the thermodynamic limit to obtain the constant-$p$ trajectories shown in Fig.~\ref{fig:1DPhaseDiagramwithp3.4linep4.5line&p5line}.

The commensurate $p=5$ trajectory is shown as the blue-gradient line labeled ``5'' in Fig.~\ref{fig:1DPhaseDiagramwithp3.4linep4.5line&p5line}, and is parametrized by
$R_b/a=3.8032+0.07847(\Delta/\Omega)+0.03372(\Delta/\Omega)^2-0.004463(\Delta/\Omega)^3$. It intersects the $\mathbb{Z}_5$ ordered region near $\Delta/\Omega\simeq 3$. Along this commensurate trajectory, the period-five melting of the Rydberg chain has the same spatial dimensionality and the same broken $\mathbb{Z}_5$ symmetry as the five-state quantum clock model. It is therefore expected to share the same universal critical behavior, namely a two-step BKT melting scenario with an intermediate critical phase separating the disordered and $\mathbb{Z}_5$ ordered phases. This is distinct from generic incommensurate cuts, where the ordered-side transition is described by a PT mechanism rather than a BKT transition. This distinction is consistent with the field-theory description, since the PT model describes a commensurate-incommensurate transition driven by a finite wave-vector mismatch and reduces to the sine-Gordon model in the commensurate limit with vanishing driving wave vector $Q=0$ \cite{LazaridesPT2009}. Thus, the ordered-side BKT transition can be viewed as the special commensurate point on the PT boundary. Therefore, an accurate determination of the constant-$p=5$ trajectory is necessary for resolving the two BKT transition points.

The $p$-state quantum clock model is defined as
\begin{equation}
    H_{\rm clock}^{(p)}
    =
    -\sum_j
    \left(\sigma_j\sigma_{j+1}^{\dagger}+\text{H.c.}\right)
    -h\sum_j
    \left(\tau_j+\text{H.c.}\right),
    \label{eq:clock_Hamiltonian}
\end{equation}
where $\sigma_j$ and $\tau_j$ are the standard clock operators satisfying $\sigma_j^p=\tau_j^p=1$ and $\sigma_j\tau_j=\omega\tau_j\sigma_j$, with $\omega=e^{2\pi i/p}$. In the $\tau$ basis, $\tau\ket{\alpha}=\omega^\alpha\ket{\alpha}$ with $\alpha=0,1,\ldots,p-1$, and $\sigma$ shifts the clock state as $\sigma\ket{\alpha}=\ket{\alpha-1}$ modulo $p$ in this convention. The large-$h$ phase is polarized by the $\tau$ field and is disordered with respect to the $\sigma$ order parameter, while the small-$h$ phase spontaneously breaks the $\mathbb{Z}_p$ symmetry and is ordered in the $\sigma$ representation. For $p\geq 5$, these two phases are separated by an intermediate critical phase bounded by two BKT transitions. In the $\tau$ representation used in Ref.~\cite{PhysRevLett.134.076501}, the field perturbation has scaling dimension $K^{\tau}$ and becomes marginal at $K^\tau_{c1}=2$, while the dual $p$-fold clock anisotropy has scaling dimension $p^2/(4K^{\tau})$ and becomes marginal at $K^\tau_{c2}=p^2/8$. These two marginality conditions give the two BKT thresholds in the $\tau$ representation.

For the Rydberg chain, the relevant order parameter is the density modulation at wave vector $2\pi/5$. The repulsive density-density interaction favors one of five translated density-wave patterns, which plays the same role as the clock-coupling term selecting one of the $\mathbb{Z}_5$-broken states in the $\sigma$ representation. We therefore convert the clock-model critical values to the $\sigma$ representation. In the convention of Ref.~\cite{PhysRevLett.134.076501}, the two dual representations satisfy $K^{\sigma}K^{\tau}=1/4$. Hence, for $p=5$, the corresponding critical values in the $\sigma$ representation are
\begin{equation}
    K_{c1}=\frac{1}{4K^\tau_{c1}}=\frac{1}{8},
    \qquad
    K_{c2}=\frac{1}{4K^\tau_{c2}}=\frac{2}{p^2}=\frac{2}{25}.
    \label{eq:p5_Kc_values}
\end{equation}
Here $K_{c1}=1/8$ identifies the BKT transition between the disordered and floating phases, where the Rabi term generates the dual vertex operator that becomes marginal and gaps the TLL on the disordered side. The second value, $K_{c2}=2/25$ identifies the BKT transition between the floating and $\mathbb{Z}_5$ ordered phases, where the fivefold anisotropy associated with domain-wall excitations becomes marginal and pins the density wave. This conversion is illustrated in Fig.~\ref{fig:p-clocksigmaBasiscrosscap}.

We further compare the Rydberg chain with the five-state clock model using the rescaled energy gap $N\Delta E$. For the clock model in the ordered phase, the lowest five states become nearly degenerate in a finite system. Therefore, extracting the physical excitation gap without lifting this degeneracy would require targeting at least six low-energy states in DMRG. To avoid this difficulty, we add the boundary pinning term
\begin{equation}
    H_{\rm pin}
    =
    -\Lambda
    \left(\sigma_1+\sigma_N+\text{H.c.}\right)
    \label{eq:clock_pinning_term}
\end{equation}
to Eq.~\eqref{eq:clock_Hamiltonian}. This term selects one of the five ordered states and separates the remaining symmetry-related states from the ground state. With this pinning field, the physical excitation gap can be extracted by targeting only the two lowest states. We set $\Lambda=1$ in the calculations. Figure~\ref{fig:NdEvsDelta-clock5}(a) shows $N\Delta E$ for the pinned clock model as a function of $h$. Two crossings are visible across the floating regime, consistent with two BKT transitions. The Rydberg chain displays the same qualitative structure. As shown in Fig.~\ref{fig:NdEvsDelta-clock5}(b), $N\Delta E$ along the commensurate $k=2\pi/5$ line also shows two crossings within a narrow parameter window. This double-crossing structure indicates that the low-energy physics of the Rydberg chain along the commensurate $p=5$ line is described by the same two-step BKT scenario as the five-state clock model.

The Binder cumulant provides an additional diagnostic of the BKT behavior along the commensurate $p=5$ line. Following our previous work \cite{PhysRevB.111.165154}, we include the logarithmic correction associated with the essential singularity of a BKT transition and analyze the modified cumulant $U_4'=U_4(1+A/\ln N)$. Near a BKT transition, the correlation length diverges as $\xi\sim \exp(b/\sqrt{\delta})$, where $\delta=(\Delta/\Omega)_c-\Delta/\Omega$ measures the distance from the critical point and $b$ is nonuniversal. Since $U_4'$ is dimensionless, it is expected to collapse onto a universal scaling function of $N/\xi$, or equivalently of $x_N=\ln(N/\xi)=\ln N-b/\sqrt{\delta}$. As shown in Fig.~\ref{fig:KTU4forp5line}, the data for $N=601,651,701,751$ collapse well with $(\Delta/\Omega)_c=2.8308$, $A=0.03$, and $b=0.05$. The inset shows the raw Binder cumulant $U_4$, which exhibits only a single crossing in this parameter window. This is consistent with the fact that the two BKT transitions are very close, so the Binder cumulant of the $\mathbb{Z}_5$ order parameter mainly captures a single effective crossing rather than resolving the two transition points separately. Thus, the Binder-cumulant analysis supports the BKT nature of the melting process, while the two nearby BKT points are resolved more clearly from the rescaled energy gap and Luttinger parameter discussed below.

\begin{figure*}[htp]
\centering    
\includegraphics[width=1\linewidth]{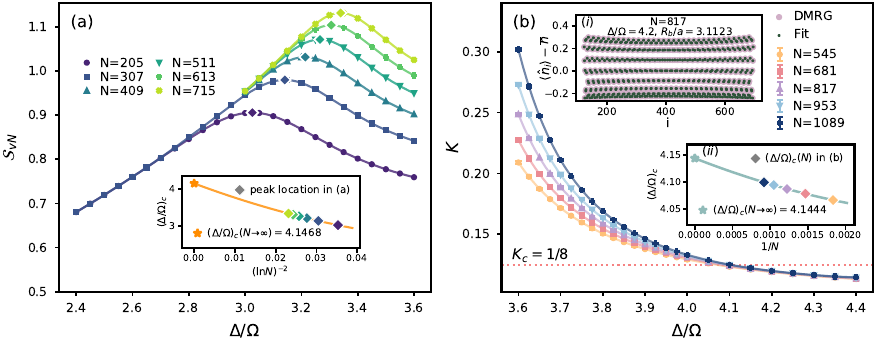}
\caption{Entanglement entropy $\mathcal{S}_{\rm vN}$ and Luttinger parameter $K$ across the BKT transition between the disordered and floating phases along the incommensurate constant-$p=3.4$ line. (a) $\mathcal{S}_{\rm vN}$ as a function of $\Delta/\Omega$ for different system sizes. The diamond symbols mark the maxima of the entanglement entropy, which determine the finite-size BKT pseudocritical points $(\Delta/\Omega)_c(N)$. The inset shows the extrapolation of $(\Delta/\Omega)_c(N)$, using the fitting function $f(x)=4.1468-40.8966x^2+258.9172x^4$ with $x=1/\ln N$. (b) Luttinger parameter $K$ extracted from Friedel oscillations as a function of $\Delta/\Omega$ for different system sizes. The intersections between the finite-size $K$ curves and the BKT critical value $K_c=1/8$ determine $(\Delta/\Omega)_c(N)$. Inset (i) shows a representative Friedel-oscillation fit for $N=817$, $\Delta/\Omega=4.2$, and $R_b/a=3.1123$, giving $K=0.1195$. Inset (ii) shows the extrapolation of $(\Delta/\Omega)_c(N)$, using the fitting function $f(x)=6769.5793x^2-54.7077x+4.1444$ with $x=1/N$.}
\label{fig:SvN_Friedel_BKTRydChain3.4pline&crosscap}
\end{figure*}

We next extract the Luttinger parameter $K$ along the constant-$p=5$ line using both Friedel oscillations and the crosscap overlap. Figure~\ref{fig:1Dp5lineFriedelOscillation&crosscap}(a) shows the finite-size $K$ obtained from Friedel oscillations as a function of $\Delta/\Omega$. The fits use the multi-harmonic ansatz in Eq.~\eqref{eq:harmonicFrideloscillation}, with harmonics included up to $m=3$. We use system sizes from $N=601$ to $1201$ with increment $120$. In the Friedel-oscillation fits, the wave vector is fixed to $q=2\pi/5$, and $15\%$ of the data are discarded at each edge. A representative fit for $N=1081$ at $(\Delta/\Omega,R_b/a)=(2.835,4.195)$ is shown in the inset of Fig.~\ref{fig:1Dp5lineFriedelOscillation&crosscap}(a). The fitted profile agrees very well with the DMRG density data, giving $K=0.0392$. The intersections between the finite-size $K$ curves and the critical values $K_{c1}=1/8$ and $K_{c2}=2/25$ determine the corresponding pseudocritical BKT points $(\Delta/\Omega)_c(N)$.

The pseudocritical points $(\Delta/\Omega)_c(N)$ obtained from Friedel oscillations are plotted as a function of $1/N$ in Fig.~\ref{fig:1Dp5lineFriedelOscillation&crosscap}(c). We fit them using second-degree polynomials in $1/N$ and extrapolate to the thermodynamic limit. The fitting functions are $f_1(1/N)=-330.2853(1/N)^2+1.1576(1/N)+2.8216$ for the left BKT point between the disordered and floating phases, and $f_2(1/N)=-206.5301(1/N)^2+4.6202(1/N)+2.8234$ for the right BKT point between the floating and $\mathbb{Z}_5$ ordered phases. This gives $(\Delta/\Omega)_{c1}=2.8216$ and $(\Delta/\Omega)_{c2}=2.8234$. The two transition points are extremely close, which explains why the entanglement entropy does not show two clearly separated peaks and why the Binder-cumulant analysis captures only one effective crossing in finite systems.

We also compute $K$ using the crosscap-overlap method. Figure~\ref{fig:1Dp5lineFriedelOscillation&crosscap}(b) shows the size-dependent $K$ extracted from the crosscap overlap along the same constant-$p=5$ line. For this calculation, the system size must be even and a multiple of five, so that it is compatible with both the crosscap geometry and the period-$5$ density modulation. We use system sizes from $N=20$ to $100$ with increment $20$. The intersections of the finite-size $K$ curves with $K_{c1}=1/8$ and $K_{c2}=2/25$ again give two sets of pseudocritical BKT points. The second-degree polynomial extrapolations are shown in Fig.~\ref{fig:1Dp5lineFriedelOscillation&crosscap}(d), with fitting functions $g_1(1/N)=2.8244-4.1811(1/N)-3.5490(1/N)^2$ and $g_2(1/N)=2.8278+0.2463(1/N)+10.0673(1/N)^2$ for the two BKT points, respectively. The extrapolated transition points are $(\Delta/\Omega)_{c1}=2.8244$ and $(\Delta/\Omega)_{c2}=2.8278$. These values are slightly shifted to larger $\Delta/\Omega$ compared with the Friedel-oscillation estimates. This shift is consistent with the tendency of the crosscap method to overestimate $K$ near BKT transitions in finite systems, as also observed in Ref.~\cite{PhysRevLett.134.076501}, where it was attributed to finite-size corrections from marginal operators.

\subsubsection{Along the incommensurate lines} \label{subsec:incommensurate}


\begin{figure*}[htp]
\centering    
\includegraphics[width=1\linewidth]{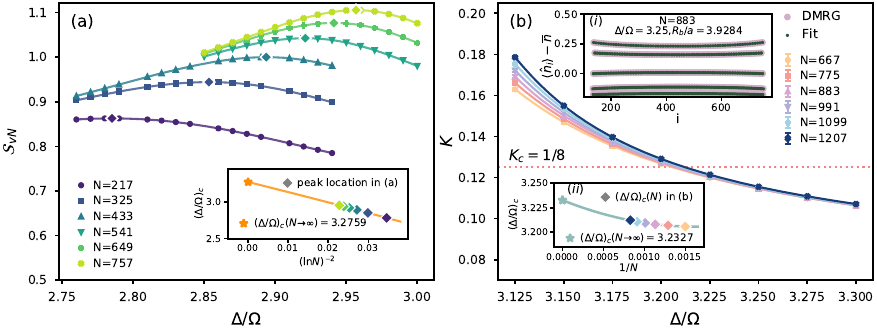}
\caption{Same as Fig.~\ref{fig:SvN_Friedel_BKTRydChain3.4pline&crosscap}, but for the incommensurate constant-$p=4.5$ line. In (a), the inset shows the extrapolation of the entanglement-entropy maxima using $f(x)=3.2759-13.6421x^2-16.3948x^4$ with $x=1/\ln N$, giving $(\Delta/\Omega)_c=3.2759$. In (b), inset (i) shows a representative Friedel-oscillation fit for $N=883$, $\Delta/\Omega=3.25$, and $R_b/a=3.9284$, giving $K=0.1147$. Inset (ii) shows the corresponding extrapolation using $f(x)=10319.7382x^2-33.3706x+3.2327$ with $x=1/N$, giving $(\Delta/\Omega)_c=3.2327$.}
\label{fig:SvN_Friedel_BKTRydChain4.5pline&crosscap}
\end{figure*}

We now turn to the incommensurate (IC) floating phases located between adjacent $\mathbb{Z}_p$ crystalline lobes. The phase diagram contains two representative IC regions, one between the $\mathbb{Z}_3$ and $\mathbb{Z}_4$ lobes with $3<p<4$, and the other between the $\mathbb{Z}_4$ and $\mathbb{Z}_5$ lobes with $4<p<5$. To extract the Luttinger parameter in the IC floating phase and locate the BKT transition between the disordered and floating phases, we mainly use Friedel oscillations. The fits use the multi-harmonic ansatz in Eq.~\eqref{eq:harmonicFrideloscillation}, with harmonics included up to $m=3$. As an independent check, we also locate the transition from the finite-size scaling of the entanglement-entropy peak.

We first consider the constant-$p=3.4$ trajectory in the IC region between the $\mathbb{Z}_3$ and $\mathbb{Z}_4$ lobes. This line is parametrized by
$R_b/a=2.4146+0.22143(\Delta/\Omega)-0.01317(\Delta/\Omega)^2$, obtained by fitting fixed-$p=3.4$ points in the thermodynamic limit on the disordered side. Figure~\ref{fig:SvN_Friedel_BKTRydChain3.4pline&crosscap}(b) shows the Luttinger parameter $K$ extracted from Friedel oscillations as a function of $\Delta/\Omega$ for different system sizes. Inset (i) shows a representative fit for $N=817$, $\Delta/\Omega=4.2$, and $R_b/a=3.1123$ inside the floating phase. The fitted profile agrees well with the DMRG density data and gives $K=0.1195<1/8$. Along this trajectory, $K$ decreases upon entering the floating phase. On the disordered side, where the density modulation decays exponentially, fitting the profile with a power-law form gives an effective $K$ that increases with system size. In contrast, inside the floating phase, the extracted $K$ values for different sizes nearly collapse. The finite-size BKT points $(\Delta/\Omega)_c(N)$ are then determined from the intersections between the $K$ curves and the critical value $K_c=1/8$. Extrapolating these pseudocritical points in $1/N$, as shown in inset (ii), gives $(\Delta/\Omega)_c=4.1444$ in the thermodynamic limit.

The entanglement entropy provides a consistent estimate of the same transition. As shown in Fig.~\ref{fig:SvN_Friedel_BKTRydChain3.4pline&crosscap}(a), $\mathcal{S}_{\rm vN}$ along the same constant-$p=3.4$ line develops a finite-size peak near the BKT transition. On the floating side of the peak, $\mathcal{S}_{\rm vN}$ increases with $N$, consistent with a critical phase. We interpolate the $\mathcal{S}_{\rm vN}$ data and determine $(\Delta/\Omega)_c(N)$ from the peak positions, marked by the diamond symbols. Extrapolating these points as a function of $(\ln N)^{-2}$ gives $(\Delta/\Omega)_c=4.1468$, in close agreement with the Friedel-oscillation estimate. This agreement supports the identification of a BKT transition between the disordered and floating phases and confirms that the multi-harmonic Friedel-oscillation analysis reliably determines the BKT boundary in the IC floating region.

We next consider the constant-$p=4.5$ trajectory between the $\mathbb{Z}_4$ and $\mathbb{Z}_5$ lobes. This line is parametrized by
$R_b/a=3.1446+0.29308(\Delta/\Omega)-0.01597(\Delta/\Omega)^2$. The results are shown in Fig.~\ref{fig:SvN_Friedel_BKTRydChain4.5pline&crosscap}. The Friedel-oscillation analysis in Fig.~\ref{fig:SvN_Friedel_BKTRydChain4.5pline&crosscap}(b) again shows that $K$ decreases from the disordered side toward the floating phase. A representative fit for $N=883$, $\Delta/\Omega=3.25$, and $R_b/a=3.9284$ gives $K=0.1147$. The intersections of the finite-size $K$ curves with $K_c=1/8$ determine the pseudocritical points, whose extrapolation in $1/N$ gives $(\Delta/\Omega)_c=3.2327$. The entanglement-entropy analysis in Fig.~\ref{fig:SvN_Friedel_BKTRydChain4.5pline&crosscap}(a) gives $(\Delta/\Omega)_c=3.2759$ from the extrapolation of the peak positions against $(\ln N)^{-2}$. The difference between the two estimates is larger than for the $p=3.4$ line, which is reasonable because BKT transitions have strong logarithmic finite-size corrections and the larger IC period effectively reduces the number of density-wave periods available in finite systems. Nevertheless, both diagnostics consistently identify the transition from the disordered phase into the IC floating phase.

These results show that the BKT boundary of the IC floating phase can be traced by the Luttinger parameter obtained from Friedel oscillations, with the entanglement-entropy peak providing an independent validation. The Friedel-oscillation method is especially useful here because the decay exponent of the leading density modulation is the Luttinger parameter $K$ in our convention. The fitted density profile therefore gives $K$ directly, and the BKT transition can be located by comparing this value with the critical threshold $K_c=1/8$. The inclusion of higher harmonics further stabilizes the extraction of $K$ in the small-$K$ regime.

\begin{figure*}[htp]
\centering    
\includegraphics[width=1\linewidth]{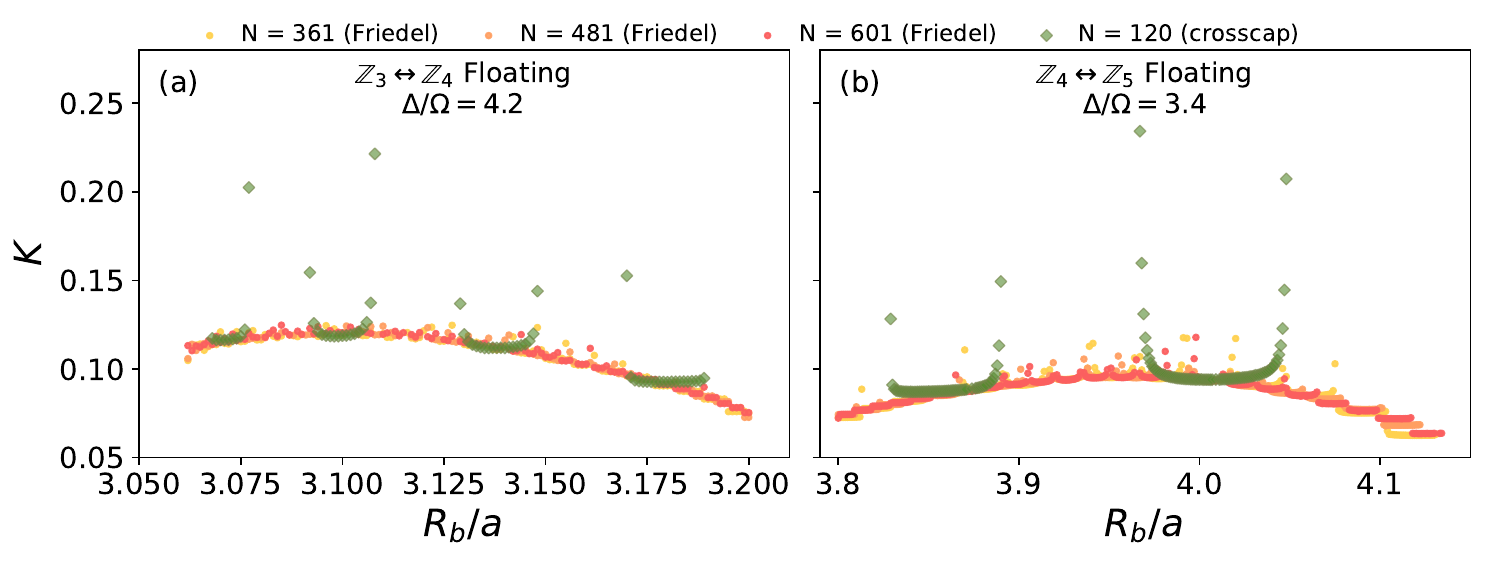}
\caption{Comparison of the Luttinger parameter $K$ extracted from Friedel oscillations and the crosscap-overlap method. (a) Vertical cut at $\Delta/\Omega=4.2$, where the floating phase lies approximately in $R_b/a\in(3,3.2)$ between the $\mathbb{Z}_3$ and $\mathbb{Z}_4$ ordered phases. The Friedel-oscillation results for different system sizes collapse onto a smooth curve, and four local minima from the crosscap-overlap method fall on the same curve. (b) Vertical cut at $\Delta/\Omega=3.4$, where the floating phase lies approximately in $R_b/a\in(3.8,4.2)$ between the $\mathbb{Z}_4$ and $\mathbb{Z}_5$ ordered phases. Two local minima from the crosscap-overlap method fall on the collapsed Friedel-oscillation curve.}
\label{fig:VcutforFriedel&crosscap}
\end{figure*}
We finally apply the crosscap-overlap method to extract $K$ in the IC floating phase. In this regime, the method is strongly constrained by finite-size commensurability. The crosscap contraction pairs site $i$ with the antipodal site $i+N/2$ and therefore requires an even system size \cite{PhysRevLett.134.076501}. For a density wave with wave vector $k$, the two paired sites have the same oscillation phase only when $kN/2$ is an integer multiple of $2\pi$. Equivalently, the periodic chain must contain an even number of complete density-wave periods. If this condition is not satisfied, the antipodal contraction pairs different phases of the density modulation and strongly suppresses the crosscap overlap. In the ideal single-mode limit, such a phase mismatch would make the overlap vanish. Since the crosscap estimator satisfies Eq.~\eqref{crosscapEq}, the compatible points give the largest overlap and therefore the smallest extracted value of $K$. These minima are expected to be closest to the physical Luttinger parameter of the single-mode floating phase in the thermodynamic limit.

This behavior is demonstrated in Fig.~\ref{fig:VcutforFriedel&crosscap}, where we compare $K$ extracted from the crosscap overlap and from Friedel oscillations along two vertical cuts inside the floating phase. For the cut at $\Delta/\Omega=4.2$, the floating phase lies approximately in $R_b/a\in(3,3.2)$ between the $\mathbb{Z}_3$ and $\mathbb{Z}_4$ ordered phases. The Friedel-oscillation results for $N=361,481,601$ collapse onto a smooth curve. For the crosscap calculation with $N=120$ under PBC, the wave-vector range corresponding to $3<p<4$ gives integer periods between 31 and 39. Among them, the even numbers $32$, $34$, $36$, and $38$ satisfy the crosscap compatibility condition. Correspondingly, the crosscap-extracted $K$ curve develops four local minima, and these minima fall on the collapsed Friedel-oscillation curve.

For the cut at $\Delta/\Omega=3.4$, the floating phase lies approximately in $R_b/a\in(3.8,4.2)$ between the $\mathbb{Z}_4$ and $\mathbb{Z}_5$ ordered phases. In this case, the compatible even numbers of complete density-wave periods in the relevant range are $26$ and $28$. The crosscap result therefore shows two local minima, which again lie on the Friedel-oscillation curve. This agreement confirms that the crosscap-overlap method remains consistent with the Friedel-oscillation extraction when the finite-size compatibility condition is satisfied. At the same time, it explains why the crosscap method is inefficient as a systematic tool for generic IC parameters in small systems, because the allowed wave vectors are discretized by the finite periodic chain, and only isolated parameter points give reliable overlaps. As the system size increases, more compatible wave vectors appear inside the IC floating phase. In the thermodynamic limit, these compatible points become dense, and the crosscap method should recover a smooth $K$ curve consistent with the Friedel-oscillation result.

\section{Conclusion} \label{sec:conclusion}

In this work, we studied the extraction of Luttinger parameters and BKT critical points in one-dimensional Rydberg atom arrays. We first benchmarked the crosscap-overlap method in two well-established models with known BKT transitions, namely the $\mathbb{Z}_3$ dual hard-core boson model and the spin-1 XY chain with single-ion anisotropy. We then revisited the five-state quantum clock model and clarified the relation between the critical Luttinger parameters in the $\tau$ and $\sigma$ representations. Since the relevant order parameter in the Rydberg chain is the density modulation, the repulsive density-density interaction selects translated density-wave patterns in the same way as the clock-coupling term in the $\sigma$ representation. The appropriate critical values are therefore $K_{c1}=1/8$ for the BKT transition between the disordered and floating phases, and $K_{c2}=2/p^2$ for the BKT transition between the floating and $\mathbb{Z}_p$ ordered phases along a commensurate trajectory.

Applying these criteria to the Rydberg chain, we showed that the commensurate $p=5$ line exhibits two successive BKT transitions. The system passes from the disordered phase into an intermediate floating phase and then into the $\mathbb{Z}_5$ ordered phase, consistent with the two-step melting scenario of the five-state clock model. The two BKT points are very close in parameter space, which explains why they are difficult to resolve from the entanglement entropy or Binder cumulant alone. They are resolved more clearly by the rescaled energy gap and by the Luttinger parameter extracted from Friedel oscillations and the crosscap overlap.

We also analyzed incommensurate floating phases between neighboring crystalline lobes. Along the constant-$p=3.4$ line between the $\mathbb{Z}_3$ and $\mathbb{Z}_4$ phases, and along the constant-$p=4.5$ line between the $\mathbb{Z}_4$ and $\mathbb{Z}_5$ phases, the transition from the disordered phase into the floating phase is consistently identified as a BKT transition. The Friedel-oscillation analysis gives the Luttinger parameter directly from the decay exponent of the leading density modulation, and the BKT point is located by the condition $K=1/8$. The finite-size scaling of the entanglement-entropy peaks provides an independent check of these phase boundaries, supporting the reliability of the multi-harmonic Friedel-oscillation method in the incommensurate regime.

From a methodological perspective, our results highlight several practical issues in extracting $K$ from finite Rydberg chains. Higher harmonics in the Friedel oscillations are important in the small-$K$ regime and must be included to obtain stable fits. It is also necessary to avoid parameter windows where the dominant wave vector changes between neighboring finite-size plateaus, since the resulting beat patterns can lead to an artificial overestimate of $K$. This effect may also help explain discrepancies in Friedel-oscillation-based estimates reported in related studies \cite{arXiv:2604.24889}. For the crosscap-overlap method, the main limitation in incommensurate floating phases is finite-size commensurability. The antipodal contraction is reliable when the periodic chain contains an even number of complete density-wave periods. Under this compatibility condition, the crosscap results agree well with the Friedel-oscillation estimates. Although this condition selects only isolated parameter points in small systems, the number of compatible wave vectors increases with system size, so the reliable crosscap estimates are expected to become dense in the incommensurate floating phase and form a smooth $K$ curve in the thermodynamic limit.

Several directions remain open. At larger blockade radii, commensurate trajectories with integer $p>5$, such as the constant-$p=6$ line, are expected to exhibit two successive BKT transitions analogous to the $p=5$ case studied here \cite{PhysRevB.100.094428,li2020}. It would also be useful to extend the present analysis to quasi-one-dimensional geometries, such as multi-leg Rydberg systems, where additional internal symmetries and density-wave structures can modify the effective clock description. Recent applications of the crosscap method to the spin-$1/2$ XYZ chain and to the commensurate regime of a Rydberg triangular-prism array \cite{losey2025calculatingluttingerliquidparameter,zuo2026groundstatephasediagramrydberg} further suggest that wave-function-based estimators of $K$ can provide a broadly useful tool for diagnosing BKT transitions in quantum simulators. In this broader context, the present work provides both a benchmark and a practical guide for applying Friedel oscillations and crosscap overlaps to interacting Rydberg systems.

\begin{acknowledgments}
We thank Hong-Hao Tu for insightful discussions. This work was supported in part by the National Natural Science Foundation of China under Grants No.~11874095 (L.-P.Y.), No.~12304172 (J.Z.), and No.~12547101, and by the Chongqing Natural Science Foundation under Grant No.~CSTB2024YCJH-KYXM0064 (J.Z.).
\end{acknowledgments}

\end{document}